\documentclass{PoS}
\usepackage{graphicx}

\def\la{\langle}
\def\ra{\rangle}
\newcommand{\be}{\begin{equation}}
\newcommand{\ee}{\end{equation}}
\newcommand{\bea}{\begin{eqnarray}}
\newcommand{\eea}{\end{eqnarray}}
\newcommand{\nn}{\nonumber}

\newcommand{\gev}{\,{\rm GeV}}   
\newcommand{\ri}{\mbox{RI-MOM}}
\newcommand{\bo}{\raise-1mm\hbox{\Large$\Box$}}    
\newcommand{\g}{\gamma}

\newcommand{\as}{\alpha_s}
\newcommand{\bk}{B_K}
\newcommand{\bkh}{\hat{B}_K}
\newcommand{\sbar}{\overline{s}}

\newcommand{\kzb}{\overline{K^0}}
\newcommand{\pkzb}{\overline{P^0}}
\newcommand{\kz}{K^0}
\newcommand{\pkz}{P^0}
\renewcommand{\vec}[1]{\mathbf{#1}}

\PoS{PoS(LAT2005)365}
\title{Kaon B-parameter with $N_F=2$ dynamical Wilson fermions}
\ShortTitle{Kaon B-parameter}
\author{SPQcdR Collaboration:}
\author{\speaker{Federico Mescia}\\
\llap{$^a$}INFN, Laboratori Nazionali di Frascati,
	    Via E. Fermi 40, I-00044 Frascati, Italy.\\
\llap{$^b$}Dip. di Fisica, Univ.
di Roma Tre, Via della Vasca Navale 84, I-00146 Roma, Italy.\\
E-mail: \email{mescia@fis.uniroma3.it}}
\author{V. Gimenez$^c$,
	    V. Lubicz$^{bd}$, G. Martinelli$^{e}$, 
S. Simula$^{d}$ 
and C. Tarantino$^{bd}$\\
\llap{$^c$} Dep. de Fisica Te\`{o}rica and IFIC, Univ. de Val\`{e}ncia, 
Dr.Moliner 50, E-46100.\\
\llap{$^d$} INFN, Sezione di Roma III,
Via della Vasca Navale 84, I-00146 Rome, Italy \\
\llap{$^e$}Dip. di Fisica, Univ. di Roma
``La Sapienza'' and INFN-Sezione di Roma, Piazzale A. Moro 2, I-00185 Roma,
Italy.}

\abstract{We present a preliminary study of the neutral kaon mixing bag 
parameter $\bk$ using two flavors of dynamical Wilson fermions. We
determine the matrix element of the relevant $\Delta S=2$ operator by
using both the conventional approach and the so called ``non-subtraction
method'', and find that the latter leads to results with smaller
uncertainties. After having implemented non-perturbative renormalization,
we study the dependence of $\bk$ on the see quark mass. At our relatively
heavy values of quark masses ($M_P/M_V \simeq 0.60 \div 0.75$) such a
dependence is found to be negligible and the results, within the
statistical accuracy, are consistent with a quenched
determination. As a preliminary result for the renormalization group
invariant parameter we quote $\bkh=1.02(25)$.}

\FullConference{XXIIIrd International Symposium on Lattice Field Theory\\
                 25-30 July 2005\\
                 Trinity College, Dublin, Ireland}

\begin{document}

\section{Preamble}
\vspace{-0.2cm}
The parameter $\bk(\mu)$, defined through
\vspace{-0.3cm}
\be
\label{eq:bkdef}
 \la\kzb|Q^{\Delta S=2}(\mu)\mid\kz\ra=\frac{8}{3}F_K^2 m_K^2\,\bk(\mu)\,,
\ee
with
\vspace{-0.3cm}
\be 
Q^{\Delta S=2}=\sbar{\g_\mu(1-\g_5)}d\;\sbar\g^\mu(1-\g_5)d 
\label{eq:qds2}
\ee
includes the long distance QCD effects in the evaluation of the indirect
CP-violating parameter $\varepsilon_K$. The theoretical uncertainty in the
determination of $\bkh$ represents at present one of the main sources of
error in the unitarity triangle analysis~\cite{Bona:2005vz}. Within the
quenched approximation the lattice estimates of $\bk$ have reached an
accuracy better than 10\%~\cite{Hashimoto:2004hn}-\cite{Dimopoulos:2004xc},
so that the quenching effect, which has been estimated to be as high as
$\pm15\%$~\cite{Sharpe:1996ih}, remains the primary source of systematic
error to be investigated. Preliminary unquenched studies are presented
in~\cite{Becirevic:2000ki}. 

In this talk we report on a exploratory calculation using $N_f=2$
degenerate flavors of dynamical fermions. With respect to previous
calculations, we simulate lighter sea quark masses on a finer lattice, but
we work with unimproved Wilson fermions. We use the Wilson plaquette
gauge action and the Wilson quark action at $\beta = 5.8$,
which corresponds to a lattice spacing $a^{-1}\simeq 3.2\gev$. The
lattice volume is $24^3\times48$. The numerical simulation is performed by
using the Hybrid Monte Carlo (HMC) algorithm~\cite{Duane}. A sample of
$50$ configurations have been generated at four different values of sea
quark masses, for which the ratio of the pseudoscalar over vector meson
masses lies in the range $M_P/M_V \simeq 0.60 \div 0.75$. Further details
on the numerical simulation can be found in~\cite{masse}. Our kaons are
made with degenerate valence quarks and we considered four values of
valence quark masses for each dynamical quark.
 
To extract $\bk$, the following three steps are needed:
\begin{itemize}
\vspace{-0.3cm}
\item
renormalization of the relevant $\Delta S=2$ operators;
\vspace{-0.3cm}
\item
extraction of the matrix elements from the appropriate correlation
functions;
\vspace{-0.3cm}
\item
determination of $\bk$ and extrapolation  to the physical quark masses. Since we use Wilson
fermions, which break chiral symmetry at ${\cal O}(a)$, cautions are
required in this latter step~\cite{Gavela:1987bd}. In particular, in
addition to the finite mixing with $dim=6$ operators, 
we have to subtract the effects of $dim\ge7$
operators at finite lattice spacing.
\vspace{-0.4cm}
\end{itemize}

\section{Renormalization and Matrix elements}
\vspace{-0.2cm}

The first ingredient in the evaluation of $\bk$ is the renormalization of 
the $Q^{\Delta S=2}$ operator of eq.~(\ref{eq:qds2}). From
eq.~(\ref{eq:bkdef}), we see that only the parity even component of
$Q^{\Delta S=2}$, namely $Q_1=\sbar{\g_\mu}d\;\sbar\g^\mu d
+\sbar{\g_\mu\g_5}d \;\sbar\g^\mu\g_5 d$, gives a non-vanishing
contribution to $\bk$. 

\begin{figure}[t]
\begin{center}
\includegraphics[width=6.5cm]{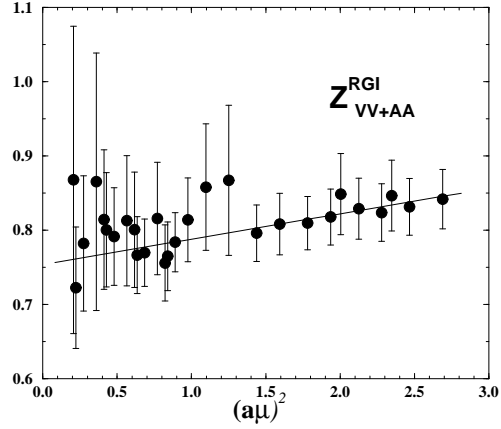}
\vskip -0.3cm
\caption{\label{fig:uno} The renormalization constant $Z^{RGI}_{VV+AA}$
obtained with the RI-MOM method. A residual ${\cal O}((a\mu)^2)$
dependence is observed. Our value of $Z^{RGI}_{VV+AA}$ in the text is given by the intercept of a linear fit in
$(a\mu)^2$ (black line).}
\vspace{-0.8cm}
\end{center}
\end{figure}

In regularizations with exact chiral symmetry, such as continuum
dimensional regularization or Ginsparg-Wilson fermions on the lattice, the
operator $Q_1$ renormalizes multiplicatively. For Wilson fermions,
instead, the renormalization pattern is more involved, and can be
expressed as
\bea
\vspace{-0.2cm}
\label{eq:q1ren} \hat Q_1(\mu) =  Z_{VV+AA}(a \mu) \biggl[ \ Q_1(a) + 
\sum_{i=2}^5 \Delta_i(a) Q_i(a) \ \biggr] \;.
\eea 
Here $Z_{VV+AA}(a \mu)$ is the multiplicative renormalization constant,
present also in formulations where chiral symmetry is preserved, while
$\Delta_{2-5}(a)$ are mixing coefficients peculiar for the Wilson 
regularization. The corresponding four-fermion operators are 
\be
\vspace{-0.2cm}
\begin{array}{ll}
Q_2(\mu)=\sbar\g_\mu d\;\sbar\g^\mu
d-\sbar\g_\mu\g_5d\;\sbar\g^\mu\g_5d\nn\;,&\quad
Q_3(\mu)=\sbar d\;\sbar d+\sbar\g_5d\;\sbar\g_5d\\
Q_4(\mu)=\sbar d\;\sbar d-\sbar\g_5d\;\sbar\g_5d\;,&\quad
Q_5(\mu)=\sbar\sigma_{\mu\nu}d\;\sbar\sigma_{\mu\nu}d.\nn 
\end{array}
\ee
In this study the renormalization constants $Z_{VV+AA}(a \mu)$ and the
mixing coefficients $\Delta_{2-5}(a)$ have been computed
non-perturbatively with the RI-MOM method~\cite{ggg}. The procedure is illustrated in
details in~\cite{z4} for the quenched case. In the unquenched case, the
situation is similar. The only additional constraint is that, in order to
get these constants in a mass-independent renormalization scheme,
a chiral extrapolation in both the valence and sea quark masses has to be
performed. At this stage, the Goldstone pole effects have been
non-perturbatively subtracted~\cite{z4}, even though their contributions
turn out to be negligible. Moreover, in order to get rid of potential
${\cal O}((a\mu)^2)$ lattice artifacts, we have performed a linear fit in
$(a\mu)^2$ in the range $[0.8,\, 2.5]$. As an example, we show this
procedure in fig.~\ref{fig:uno} for the renormalization group invariant
(RGI) constant $Z^{RGI}_{VV+AA}$. 
The scheme and scale dependent $Z_{VV+AA}(a\mu)$ and/or $\bk(\mu)$ are
related to their RGI expressions through
\be
\vspace{-0.2cm}
\bkh =\left[\as(\mu)\right]^{-{\frac{\g_0}{2\beta_0}}}\left[1 + 
\frac{\as(\mu)}{4\pi} J\right]\bk(\mu) ,
\label{eq:bk_rgi}
\ee
where $\g_0=4$ and $\beta_0=
11-2 N_f/3$. We use the $\ri$ scheme for which the NLO coefficient
$J=2.83551$ with $N_f=2$. Our 
results at $\beta=5.8$ for the renormalization constant and mixing coefficients
 are $Z^{RGI}_{VV+AA} = 0.75(6),\, \Delta_2 = -0.08(10),\,
\Delta_3 = -0.05(3),\, \Delta_4 = 0.02(4),\, \Delta_5 = 0.03(2)\,.$ 
As known from quenched studies~\cite{ggg}, the practical difficulty in
the calculation of $\bk$ with Wilson fermions is not only the
non-perturbative evaluation of the subtraction constants $\Delta_{2-5}(a)$
but also the necessity to perform it with a high level of accuracy, since 
the lattice bare matrix elements $\langle Q_{2-5}\rangle$ are orders of
magnitude larger than  $\langle Q_1\rangle$. Therefore, even though the
subtraction constants are numerically small, the net effect of the
subtraction is large.

An alternative approach  that allows one to compute the matrix
element~(\ref{eq:bkdef}) without the necessity to perform the subtraction
has been developed in~\cite{bkws}. In this procedure ``without
subtractions" one uses the hadronic chiral axial Ward identity to relate
the matrix element of the operator $Q_1$ to its parity violating
counterpart, ${\cal Q}_1 =\sbar{\g_\mu}\g_5 d\;\sbar\g^\mu d +
\sbar{\g_\mu}d\;\sbar\g^\mu\g_5 d$. The latter only renormalizes
multiplicatively with the constant $Z_{VA}(a\mu)$, which we have
calculated with the $\ri$ method determining, $Z^{RGI}_{VA}=0.80(2)$. In this way,
the problem of mixing with the other dimension-six operators is
circumvented. The price to pay is that one has to compute a four-point
correlation function where one pion operator is integrated over all
lattice space-time coordinates.

\begin{figure}[t]
\begin{center}
\includegraphics[width=6.5cm]{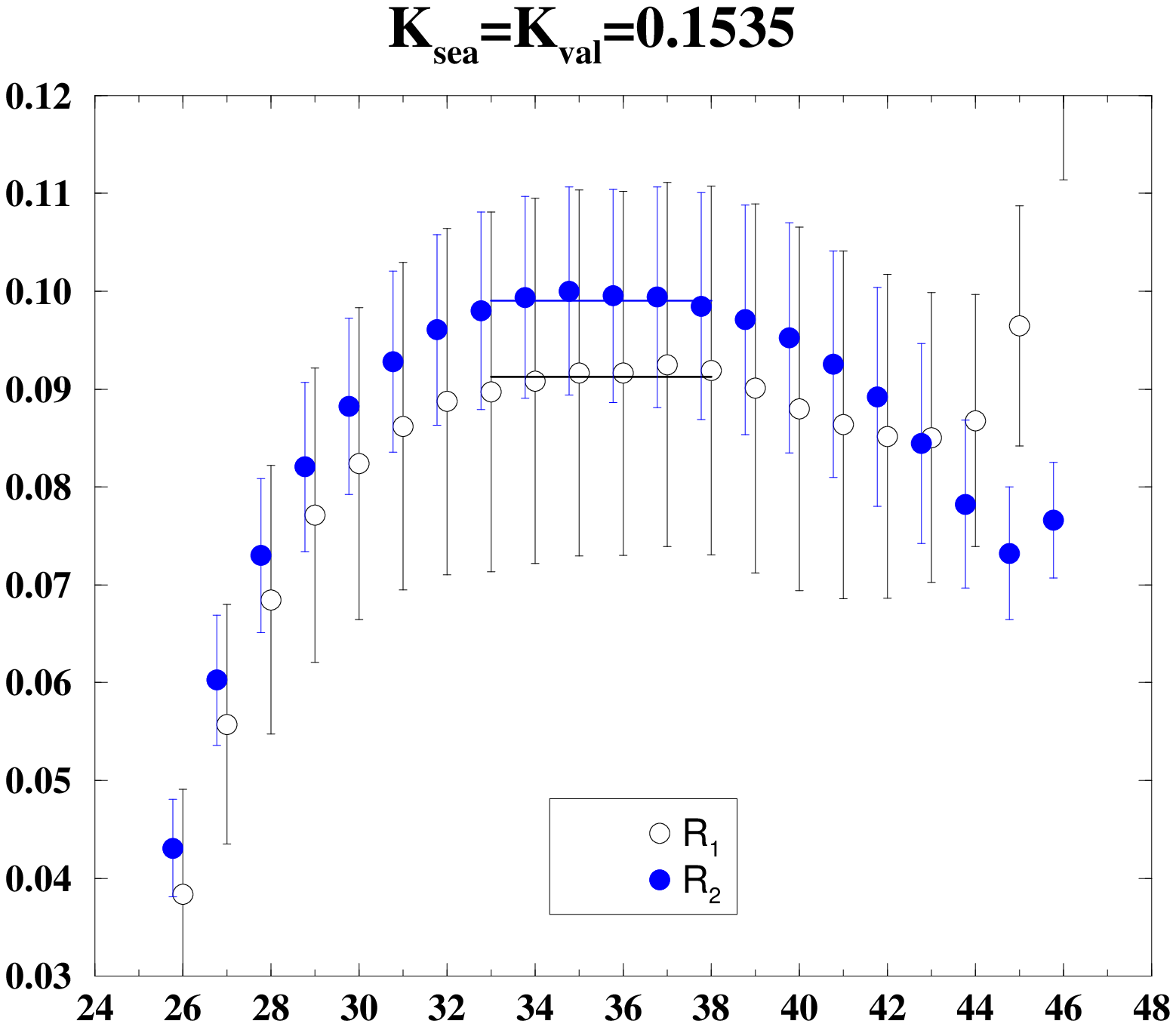}
\includegraphics[width=6.5cm]{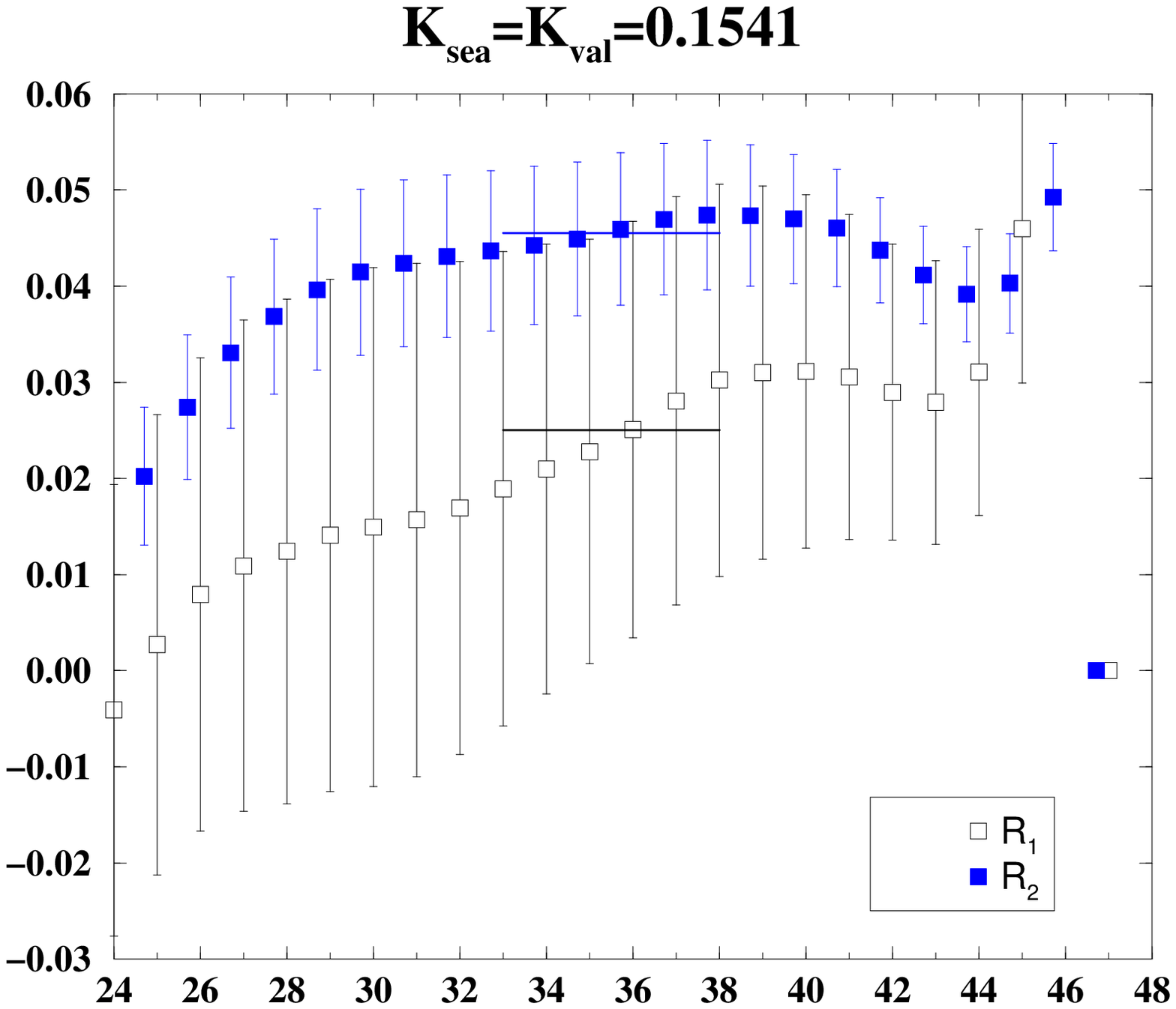}
\caption{\label{fig:plat} 
The ratios $R_1$ and $R_2$ computed with mesons
at rest. }
\end{center}
\vspace{-0.8cm}
\end{figure}

In practice, we consider the following two ratios of correlation functions
\bea
\vspace{-0.2cm}
R^{\vec p,\vec q}_1(t)&=&\frac{\sum_{\vec x\vec y}
\la P(\vec x,t) \hat Q_1(0) P(\vec y,t_f)\ra e^{i(\vec x \vec p+\vec y \vec q)}}
{\sum_{\vec x}\la P(\vec x,t)P(0) \ra e^{i\vec x \vec p} 
\sum_{\vec y}\la P(\vec y,t_f)P(0) \ra e^{i\vec y \vec q} }\\
R^{\vec p,\vec q}_2(t)&=&\frac{\sum_{\vec x\vec y z}
\la P(\vec x,t)(m^{AWI}\Pi(z)) \hat{\cal Q}_1(0) P(\vec y,t_f)\ra e^{i(\vec x \vec p+\vec y
\vec q)}}
{\sum_{\vec x}\la P(\vec x,t)P(0) \ra e^{i\vec x \vec p} 
\sum_{\vec y}\la P(\vec y,t_f)P(0) \ra e^{i\vec y \vec q} }
\eea
for the method with and without subtraction respectively. Thanks to the
Ward identity introduced in~\cite{bkws}, $R^{\vec p,\vec q}_1$ and
$R^{\vec p,\vec q}_2$ only differ by ${\cal O}(a)$ effects. At large
times $t_f,T-t\gg0$, the ratios $R_i(t)$ are both proportional to the
desired matrix element $\la\pkzb|Q^{\Delta S=2}(\mu)\mid\pkz\ra$. In our
simulation, we have chosen $t_f=14$ and $t$ in the range $[23,37]$,
whereas the momentum configurations $\{\vec p,\vec q\}$ (in units of
$2\pi/La$) are given by $\{(0,0,0),(0,0,0)\},\{(0,0,0),(1,0,0)\}$ and
$\{(1,0,0),(0,0,0)\}$. An average over momentum configurations equivalent
under hypercubic rotations is also performed. In fig.~\ref{fig:plat}, we
compare the results for $R_1$ and $R_2$ in the cases of the heaviest and 
the lightest quark masses ($m_{val}=m_{sea}$). We notice that $R_2$
suffers from rather smaller statistical fluctuations.
\vspace{-0.4cm}

\section{Chiral behavior at finite lattice spacing}
\vspace{-0.3cm}

At finite lattice spacing, the matrix elements $\la\pkzb|Q^{\Delta
S=2}(\mu)\mid\pkz\ra$ extracted from the ratios $R_i$ are no longer proportional
to $\bk$. The best approach to get rid of the ${\cal O}(a)$ contributions
coming from the mixing with $dim\ge7$ operators would be to perform a
continuum extrapolation. In the present simulation, however, we have data
only at a single value of the lattice spacing. Therefore, we rely
on the approach proposed in~\cite{Gavela:1987bd}. By calculating matrix
elements of external kaons with non-zero momentum, we introduce an
additional degree of freedom which allows us to partially remove the 
leading lattice artifacts. By writing
\be
\vspace{-0.2cm}
\label{eq:bka}
 \la\pkzb|\hat Q_1\mid\pkz,\vec q\ra_{m_{sea}}=
 \alpha_{m_{sea}}+\beta_{m_{sea}} m_P^2+ \frac{8}{3}F_P^2
m_P^2\,\gamma_{m_{sea}}\,,
\ee
\begin{figure}[t]
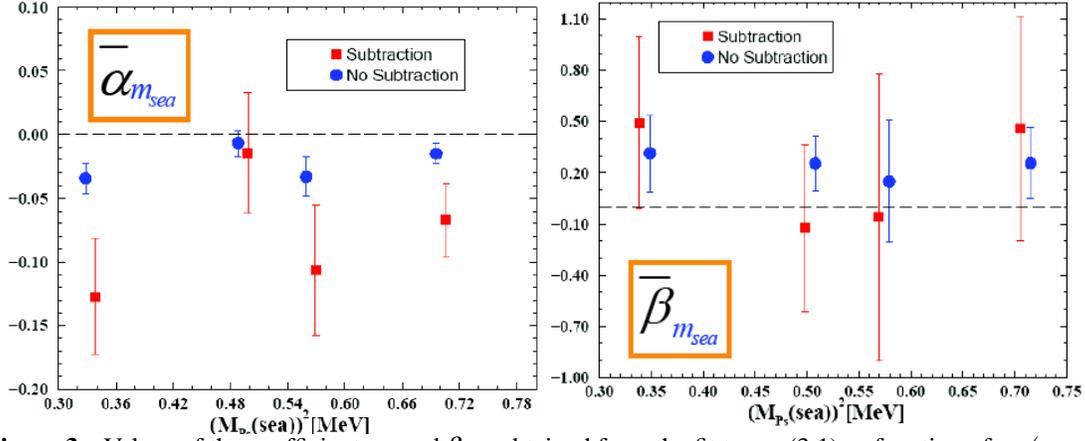

\begin{center}
\includegraphics[width=5.8cm,angle=-90]{alpha.epsi}
\includegraphics[width=5.7cm,angle=-90]{beta.epsi}
\vskip -0.3cm
\caption{\label{fig:alpha} Values of the coefficients $\alpha$ and $\beta$
as obtained from the fit to eq.~(\protect\ref{eq:bka}) as function of
$m_P(m_{sea})$.}
\end{center}
\vspace{-0.8cm}
\end{figure}
one finds that the physical contribution to $\bkh$, at each $m_{sea}$, is given by the
coefficient $\gamma$, while $\alpha$ and $\beta$ parameterizes pure
lattice artifacts. The results of the fit for the coefficients $\alpha$
and $\beta$ are shown in fig.~(\ref{fig:alpha}). These coefficients turn
to be sizable and their contribution in the extraction of the parameter
$\gamma$ cannot be neglected. It is also interesting to notice that the
effect of these coefficients is less relevant in the method without
subtraction (blue points in fig.~(\ref{fig:alpha})). 

\begin{figure}[b]
\vspace{-0.4cm}
\begin{center}
\includegraphics[width=6.5cm]{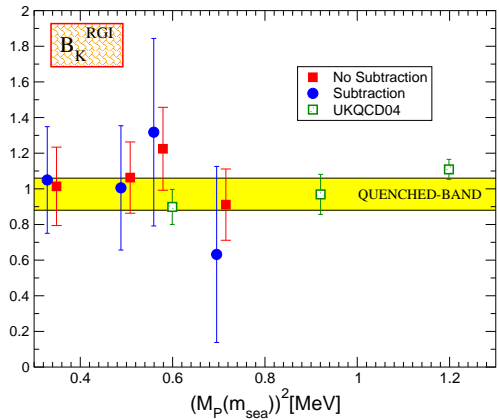}
\caption{\label{fig:scen} Values of $\bkh$, as obtained from the two 
methods with and without subtraction, as a function of the pseudoscalar
meson mass squared. The results obtained by the UKQCD
Collaboration~\cite{Becirevic:2000ki} and the band corresponding to the
quenched estimate are also shown for comparison.}
\end{center}
\end{figure}

We finally plot in fig.~\ref{fig:scen} $\bkh(m_{sea})=\gamma_{sea}$,
as a function of the square of the pseudoscalar meson mass
$m_P(m_{sea})$. We observe that the method without-subtractions
(red points) looks more promising, by suffering from smaller
uncertainties. We also compare in the plot our results with those
obtained by the UKQCD Collaboration~\cite{Becirevic:2000ki} by using
${\cal O}(a)$-improved Clover fermions but at a larger value of the
lattice spacing. The two sets of results are very well compatible. 
In order to obtain the physical value of $\bkh$, an extrapolation to
$m_P(m_{sea})=0$ should be performed. Within the errors, however, we do not see any
significant dependence on the sea quark mass and our results, at the
simulated values of quark masses, are compatible with the quenched
estimate (yellow band in fig.~\ref{fig:scen}), 
$\bkh=0.97(9)$ obtained by performing a quenched simulation on a lattice with
similar size and resolution. As a
preliminary result, we quote the value of $\bkh$ obtained from a constant
fit to the  points obtained with the method without subtraction, namely
\be
\vspace{-0.4cm}
\bkh=1.02(25)
\label{eq:bk_res}
\ee
In order to reduce the uncertainties, we plan to include in the analysis more external
momenta by exploiting  the $\theta$-boundary
condition method of ref.~\cite{tantalo}.
\vspace{-0.5cm}

\section*{Acknowledgments}
\vspace{-0.3cm}
F.~Mescia thanks the INFN-Rome section for financial support.
The work by F.Mescia was partially supported by IHP-RTN,
EC contract No.\ HPRN-CT-2002-00311 (EURIDICE).
The work by V.G. has been funded by MCyT, Plan Nacional I+D+I
(Spain) under the Grant BFM2002-00568.

\end{document}